\documentclass[twocolumn, twocolappendix]{aastex7}

\usepackage{amsmath}
\usepackage{amssymb}
\usepackage{bm}
\usepackage{xcolor}

\newcommand{\chevrons}[1]{\left\langle#1\right\rangle}
\newcommand{\chevronsi}[1]{\langle#1\rangle}
\newcommand{\abs}[1]{\left|#1\right|}

\newcommand{\ujb}{\,$\mu$Jy\,beam$^{-1}$}

\graphicspath{{./}{figs/}}
\begin{document}

\title{IXPE Polarizations of the Lighthouse Pulsar, Trail, and Filament}

\author[0000-0002-6401-778X]{Jack T. Dinsmore}
\email{jtd@stanford.edu}
\affiliation{Department of Physics, Stanford University, Stanford CA 94305}
\affiliation{Kavli Institute for Particle Astrophysics and Cosmology, Stanford University, Stanford CA 94305}

\author[0000-0001-6711-3286]{Roger W. Romani}
\email{rwr@stanford.edu}
\affiliation{Department of Physics, Stanford University, Stanford CA 94305}
\affiliation{Kavli Institute for Particle Astrophysics and Cosmology, Stanford University, Stanford CA 94305}

\author[0000-0002-2096-6051]{S. Zhang}
\affiliation{Department of Physics, The University of Hong Kong, Pokfulam, Hong Kong}
\affiliation{Hong Kong Institute for Astronomy and Astrophysics, The University of Hong Kong, Pokfulam, Hong Kong}
\email{}

\author[0000-0002-5847-2612]{C.-Y. Ng}
\affiliation{Department of Physics, The University of Hong Kong, Pokfulam, Hong Kong}
\affiliation{Hong Kong Institute for Astronomy and Astrophysics, The University of Hong Kong, Pokfulam, Hong Kong}
\email{ncy@astro.physics.hku.hk}

\author[0000-0002-8665-0105]{Stefano Silvestri}
\affiliation{Istituto Nazionale di Fisica Nucleare, Sezione di Pisa, Largo B. Pontecorvo 3, 56127 Pisa, Italy}
\email{stefano.silvestri@pi.infn.it}

\author[0000-0002-6447-4251]{Oleg Kargaltsev}
\affiliation{Department of Physics, The George Washington University, 725 21st St. NW, Washington, DC 20052}
\email{kargaltsev@email.gwu.edu}

\author[0000-0002-8848-1392]{Niccol\`{o} Bucciantini}
\affiliation{INAF Osservatorio Astrofisico di Arcetri, Largo Enrico Fermi 5, 50125 Firenze, Italy}
\affiliation{Dipartimento di Fisica e Astronomia, Universit\`{a} degli Studi di Firenze, Via Sansone 1, 50019 Sesto Fiorentino (FI), Italy}
\affiliation{Istituto Nazionale di Fisica Nucleare, Sezione di Firenze, Via Sansone 1, 50019 Sesto Fiorentino (FI), Italy}
\email{niccolo.bucciantini@inaf.it}

\author[0000-0002-3638-0637]{Philip Kaaret}
\affiliation{NASA Marshall Space Flight Center, Huntsville, AL 35812, USA}
\email{philip.kaaret@nasa.gov}

\author[0000-0001-6395-2066]{Josephine Wong}
\affiliation{Department of Physics, Stanford University, Stanford CA 94305}
\affiliation{Kavli Institute for Particle Astrophysics and Cosmology, Stanford University, Stanford CA 94305}
\email{joswong@stanford.edu}

\author[0000-0002-6986-6756]{Patrick Slane}
\affiliation{Center for Astrophysics | Harvard \& Smithsonian, 60 Garden Street, Cambridge, MA 02138, USA}
\email{pslane@cfa.harvard.edu}

\author[0000-0002-7781-4104]{Paolo Soffitta}
\affiliation{INAF Istituto di Astrofisica e Planetologia Spaziali, Via del Fosso del Cavaliere 100, 00133 Roma, Italy}
\email{paolo.soffitta@inaf.it}

\author[0000-0002-5270-4240]{Martin C. Weisskopf}
\affiliation{NASA Marshall Space Flight Center, Huntsville, AL 35812, USA}
\email{martin.c.weisskopf@nasa.gov}

\begin{abstract}
The Lighthouse pulsar (PSR J1101$-$6101) sports a bright X-ray trail and filament. The synchrotron emission from both structures is expected to be polarized, with electric vector position angle (EVPA) perpendicular to the magnetic field direction and polarization degree (PD) indicating the local degree of magnetic turbulence. We present a 1 megasecond Imaging X-ray Polarimetry Explorer (IXPE) observation of the Lighthouse complex. At the 99\% confidence level, we detect the filament polarization with PD $55\pm18\%$ and EVPA indicating a magnetic field parallel to the filament axis. The large PD implies a turbulent magnetic field weaker than the background field, in conflict with some existing models. We also detect polarization from the pulsar and trail. The trail's X-ray polarization is nearly orthogonal to the radio polarization, suggesting spatial separation between the X-ray- and radio-emitting leptons. The pulsar polarization is well-fit by the rotating vector model.
\end{abstract}


\begin{figure*}
  \centering
  \includegraphics[width=\linewidth]{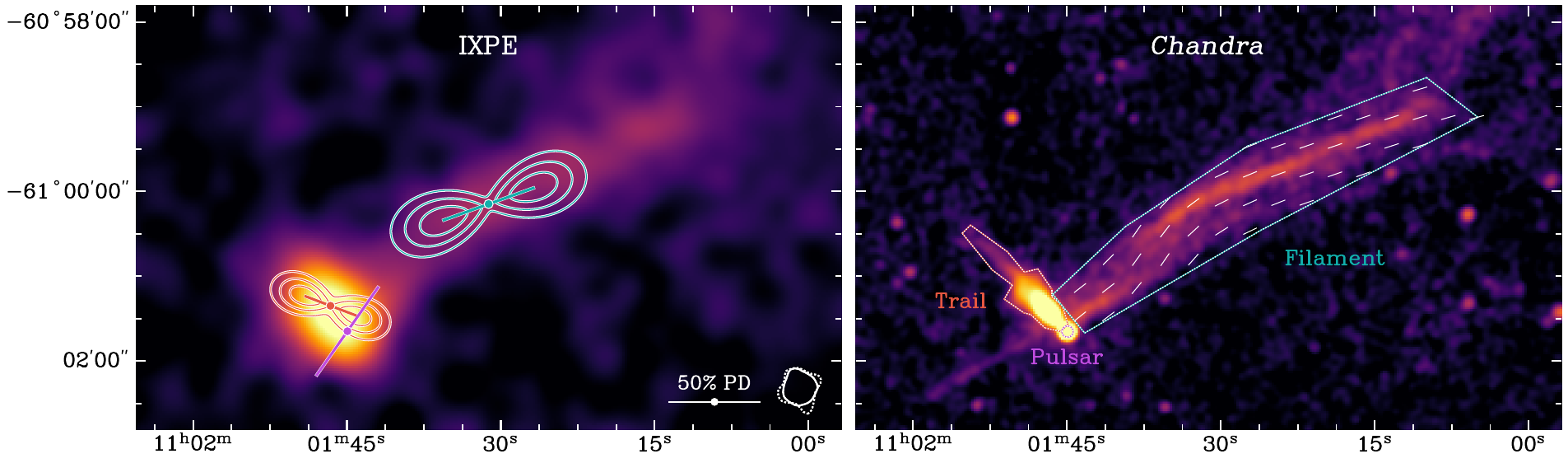}
  \caption{The Lighthouse pulsar, trail, and filament as viewed by IXPE (\textit{left}) and \textit{Chandra} (\textit{right}) at $2-8$ keV. Best-fit results and 68, 95, 99.7\% confidence intervals for the projected magnetic field position angle are shown in the IXPE image, corresponding to the regions outlined in the \textit{Chandra} image. For the pulsar, a bar shows the projected spin axis, with length scaled to the phase average RVM PD. Our model for the filament-following magnetic field, which the data prefer over a constant PA model, is shown as white lines in the \textit{Chandra} image. The PSF half light contours for DU1 (solid) and DU3 (dotted) are shown in the IXPE image.}
  \label{fig:lighthouse}
\end{figure*}

\section{Introduction} \label{sec:intro}
PSR J1101$-$6101 (hereafter J1101 or ``Lighthouse'') is a young, energetic, rotation-powered pulsar \citep[spin down age $\tau_c = P / (2\dot P) = 63$ kyr, spin down power $\dot E = 1.6 \times 10^{36}$ erg s$^{-1}$, and period $P=63$ ms,][]{halpern2014discovery}, rapidly moving at 990 km s$^{-1}$ \citep{dinsmore2026chandra}. J1101 generates two extended X-ray structures. A bright X-ray trail extends $\sim 1'$ behind the pulsar, pointing to its birth supernova remnant (SNR) MSH 11$-$61A \citep{pavan2011igr}. Trails occur when the wind from a rather young, supersonic pulsar shocks against the interstellar medium (ISM), creating a bow shock ahead of the pulsar and a trail in the turbulent wake behind (see \cite{kargaltsev2017pulsar} for a review).
The second structure is a rare X-ray filament, extending for $\sim 5'$ orthogonal to the trail axis, persisting several parsecs into the interstellar medium \citep[ISM;][]{pavan2014long} and accompanied by a fainter opposing ``anti-filament'' (see Fig.~\ref{fig:lighthouse}, right). X-ray filaments---of which Lighthouse is the brightest \citep{dinsmore2024catalog}---are characterized by their misalignment with the pulsar velocity and are proposed to represent cosmic ray (CR) leptons escaping the bow shock onto ISM fields \citep{bandiera2008on}.
Both structures emit synchrotron X-rays up to 25 keV \citep{klingler2023nustar}, but only the trail is visible at radio wavelengths, where the filament and pulsar are absent \citep{pavan2014long}.

Central to the physics of trails and filaments is the magnetic field geometry, which is accessible through polarization observations. Synchrotron emission has electric vector position angle (EVPA) perpendicular to the background magnetic field $\bm B_0$, and maximum polarization degree (PD)
\begin{equation}
\Pi_0 = \frac{\Gamma}{\Gamma + 2/3}
\label{eqn:max-pd}
\end{equation}
for power-law particle energy distributions, where $\Gamma$ is the spectrum photon index ($dN / dE_\gamma = E_\gamma^{-\Gamma}$). Turbulence or macroscopic variation in field orientation along the line-of-sight depolarize the emission, as can internal Faraday rotation at radio wavelengths.

Magnetohydrodynamic simulations depict a complex magnetic field geometry in pulsar trails \citep{olmi2019full}, with an average magnetic field depending on the pulsar spin axis and wind asymmetry. Observations also show variable geometries, with radio polarization of the Mouse \citep{yusef2005radio} and PSR J1509$-$5850 \citep{ng2010radio} showing fields parallel and perpendicular to the trail axis, respectively. No X-ray polarization observations of trails have been published to date.

If filaments indeed represent CRs propagating along ISM field lines, the background magnetic field $\bm B_0$ should align with the filament axis. Radio polarizations are not accessible, since filaments are radio-quiet, but this field geometry has been confirmed at X-ray wavelengths for the filament candidate G0.13$-$0.11 \citep{churazov2024pulsar} and via starlight polarization observations of the Guitar filament \citep{dinsmore2025starlight}. Magnetic turbulence traps filament particles, but the details are still unclear. \cite{bykov2017pulsar} and \cite{olmi2024nature} have suggested that the non-resonant streaming instability grows the filament's turbulence until saturation, when the turbulent field strength is much greater than the underlying field ($\Delta B \gg B_0$). It is also possible that the turbulence growth cuts off earlier, when the pulsar's rapid motion carries the CR injection site to new field lines \citep{dinsmore2026physical}. This earlier cutoff could allow steady-state turbulence with $\Delta B < B_0$. X-ray polarization observations can discriminate between these models.

We present a 950 kilosecond Imaging X-ray Polarimetry Explorer (IXPE) observation of Lighthouse. The data are analyzed using new techniques that substantially improve the polarization precision, as reviewed in \S\ref{sec:data}. Polarization results are discussed in \S\ref{sec:results}. We contrast our results with other analysis methods in \S\ref{sec:variations} and conclude with \S\ref{sec:conclusion}.

\section{Analysis methods} \label{sec:data}
IXPE observed Lighthouse from May 30 to June 17, 2025, for 950 ks (observation ID 04001301). Due to an anomaly with detector unit (DU) 2 that occurred on 14 April 2025, we are unable to use the DU2 data at the current time and performed the analysis using DU1 and DU3 only. EVPAs, positions, and polarization weights are reconstructed using the NN developed by \cite{peirson2021deep}. Although the NN also estimates energies, gain variations since launch render the Moments energies more reliable, so we use these. Eventually DU2 data with revised gain may become available as well. We also calculate ``particle characters'' using the NN designed by \cite{dinsmore2025advanced}, which allow a weighted removal of background particles. Spurious modulation is removed using the rescaled maps reported in \cite{dinsmore2025advanced}. Boom drift was corrected using the \texttt{ixpeboomdriftcorr} tool from \texttt{HEAsoft} version 6.33.2 \citep{nasa2014HEAsoft} and default parameters. After restricting to valid good-time intervals, we cut to 2$-$8 keV data within 247$''$ of the aim point to avoid serious vignetting.

Several bright flares pollute our data, particularly in DU3. We can use the NN-determined particle characters to determine whether flares are composed of photons or particles. Most are photon-dominated, likely due to solar activity as the exposure was taken near solar maximum. Some short, particle-dominated flares occur before entrance into the South Atlantic Anomaly (SAA). To cut flares, we bin the $2-8$ keV DU1 and DU3 data from a large background region into 2.55 minute time bins. The distribution of bin fluxes is approximately Gaussian at low count rates, breaking to an exponential tail at high count rates. We set an event rate threshold at the break rate and cut all events in bins with larger rates. This process removes 2.4\% of the total exposure.

\subsection{Fit Method}

The commonly employed \texttt{PCUBE} analysis uses polarization extraction \citep{baldini2022ixpeobssim} from events reconstructed with the ``Moments'' pipeline. While this method is an excellent tool to extract polarizations from bright sources, it does not take full advantage of IXPE's measurements. Maximum likelihood techniques can substantially improve uncertainties \citep{peirson2021deep, marshall2021multiband, gonzalez2023unbinned, ravi2025whats}, as does the neural net (NN) reduction of the raw event images we employ \citep{peirson2021deep, cibrario2025mitigating}. Advanced weighting techniques can help separate the pulsar from the trail by position \citep{dinsmore2025advanced}, phase \citep{wong2023improved}, and energy \citep{vianello2015multi} and remove polluting CRs \citep{dimarco2023handling, dinsmore2025advanced}. Our Lighthouse analysis uses the \texttt{LeakageLib} pipeline \citep{dinsmore2024leakagelib}, which combines all these techniques. This pipeline was validated by \cite{dinsmore2025advanced} on simulated and real observations.

\texttt{LeakageLib} extracts best-fit polarizations and Gaussian uncertainties by maximizing the likelihood. This allows us to test the significance of our results by calculating the delta log-likelihood $\Delta \ln L$ between the best-fit model and the appropriate null hypothesis. By Wilks' theorem, $-2\Delta \ln L$ is $\chi^2$-distributed with $k$ degrees of freedom (DoF), where $k$ is the DoF difference between the model and null hypothesis (see e.g.~\cite{algeri2020searching} for a review of Wilks' theorem and its drawbacks). Hence, the $p$-values are $p = \hat \Gamma(k/2, -\Delta \ln L)$, where $\hat \Gamma$ is the normalized upper incomplete $\Gamma$-function.

\subsection{Source Models}
To extract polarizations, we identify the astrophysical sources (the pulsar, trail, and filament) and background sources, and we simultaneously fit for the polarization of each to the data. Our spatial weights require a high-resolution source flux model and IXPE point-spread function (PSF) models for each DU. We use \textit{Chandra} observations as the flux model for the astrophysical sources (\S\ref{sec:images}), uniform distributions for the background sources, and the detailed PSFs measured by \cite{dinsmore2024polarization}. We phase-weight by the IXPE pulsar light curve (\S\ref{sec:light-curve}) and spectral-weight with spectra measured from the \textit{Chandra} data (\S\ref{sec:spectrum}).

Some particle tracks are elongated, so that reconstruction algorithms assign an EVPA $\psi$ parallel to the particle axis. The fits we report in \S\ref{sec:results} for the filament and pulsar/trail polarizations show a preference for an asymmetric distribution of $\psi$ for particles, with significance ranging from $0.9-2.2\sigma$ depending on which source is being fit, and whether NN or Moments reconstructed data are used. This has also been seen for other IXPE data sets \citep[e.g.][]{sullivan2026xray}. Though this result is not significant, modeling it eliminates a systematic that could otherwise pollute our measurement of the source polarization. We therefore treat the particle track reconstructed EVPA with a probability density function (PDF) $P(\psi) =  [1+\Pi_\mathrm{part} \cos (2(\psi - \Psi_\mathrm{part}))]/(2\pi)$ for free parameters $\Pi_\mathrm{part}$ and $\Psi_\mathrm{part}$. Background fits give $\Pi_\mathrm{part} \sim 2\%$. There is a $1.2-2.7\sigma$ preference for different parameters in each detector, so we model DU1 and DU3 independently. The photon background is also polarized to the $\sim 10\%$ level ($3.0-4.1\sigma$) and is polarized differently between detectors ($1.8-3.5\sigma$), so we separate the photon background models as well. \cite{bucciantini2025polarized} similarly saw increased flaring and photon background polarization in DU3 compared to DU1.

\subsection{Model Image Extraction} \label{sec:images}
The \textit{Chandra} X-ray archive contains 463\,ks of ACIS imaging of Lighthouse (the observations employed are listed in~\dataset[DOI: 10.25574/cdc.576]{https://doi.org/10.25574/cdc.576}). After subtracting the background flux using a large region to the south, we convert these observations to the IXPE band with the ratio of the Auxiliary Response Functions (ARF). Each $2-8$ keV \textit{Chandra} event is weighted by the ratio of the IXPE ARF (extracted using \texttt{ixpecalcarf}) to the \textit{Chandra} ARF (extracted using \texttt{mkwarf} for the relevant event file and region). The standard \texttt{ciao} and \texttt{HEAsoft} tools were used. This leaves a well-resolved, unpolarized map of the field. We divide the image into pulsar, trail, and filament components to be used as source flux models for the spatial weights (see regions in Fig.~\ref{fig:lighthouse}).

For the spatial weights to be effective, the model images must be well aligned with the IXPE data. To determine the relative offset, we convolved the stacked trail and pulsar \textit{Chandra} images with the IXPE PSFs to produce a model IXPE image of the nebula. We fitted this image to the binned IXPE data by minimizing the $\chi^2$ statistic, revealing $\sim 1''$ shifts from the reported aim point. The best-fit $\chi^2$ per degree of freedom was satisfactory, valued at $475/436$. No additional blur was required to match the data, implying that the boom drift was well-corrected by \texttt{ixpeboomdriftcorr}.

\subsection{Light Curve Extraction} \label{sec:light-curve}
\begin{figure}
  \centering
  \includegraphics[width=\linewidth]{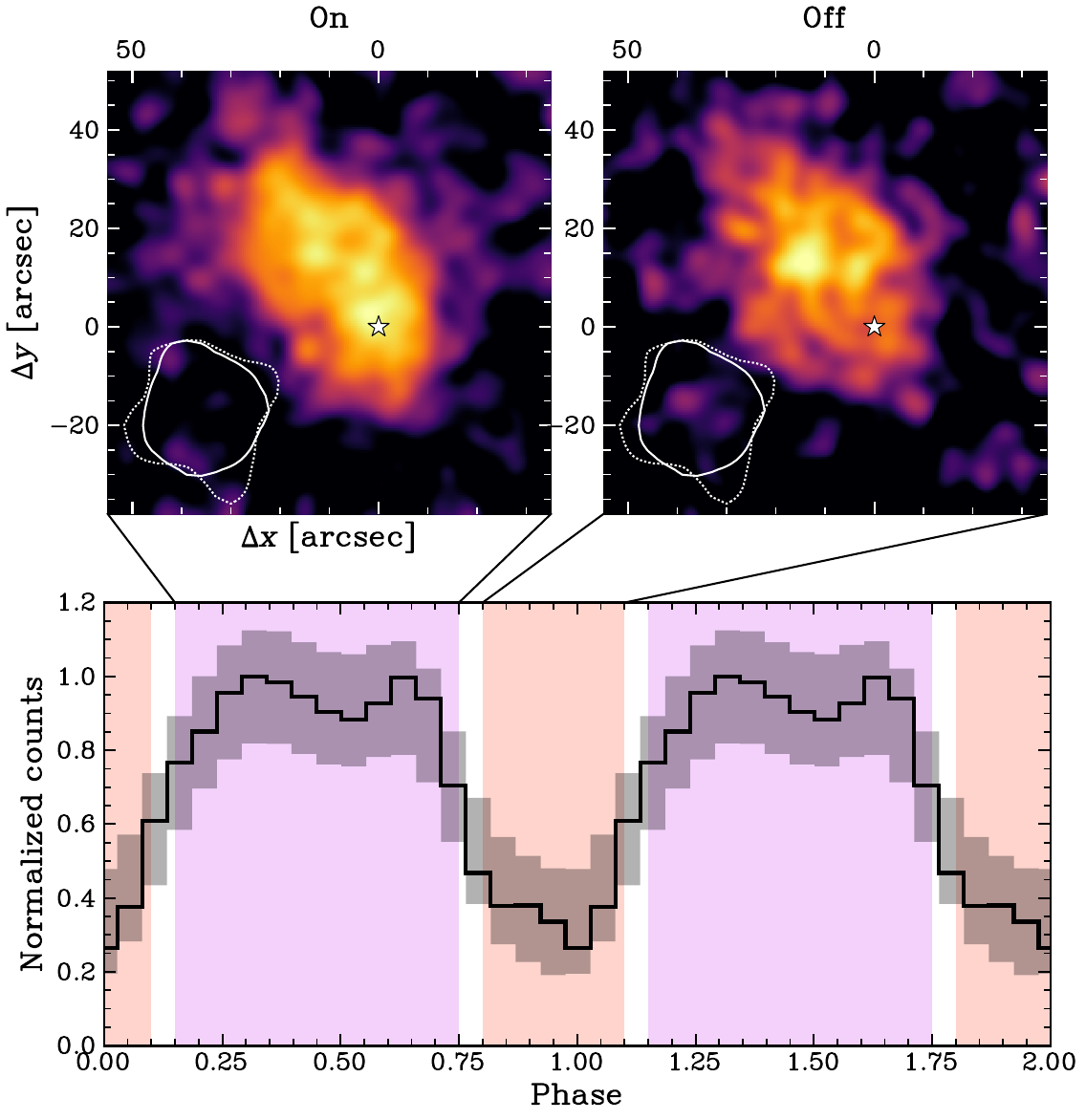}
  \caption{\textit{Top}: IXPE image of the pulsar and trail during ``on'' and ``off'' phases marked in the bottom panel. The DU1 (solid), and DU3 (dotted) PSF half light contours are shown in the lower left. The shift of the center of light in the off phase shows that the pulsar is partially resolved, so that the trail and pulsar polarizations are separable. \textit{Bottom:} Lighthouse's IXPE light curve (smoothed), with spatial and particle weights delivering improved significance.}
  \label{fig:lc}
\end{figure}

We phase the IXPE data with the \texttt{PINT} software \citep{luo2021pint, susobhanan2024pint} using the \textit{NuSTAR} / \textit{NICER} ephemeris found by \cite{ho2022timing}. A clear pulse is detected in $1-8$\,keV data using a simple extraction aperture, though with low signal to noise and a phase offset. Weighting each event by the IXPE PSFs and the particle character approximately doubles the significance of pulsations in the extracted light curve, and the result is shown in Fig.~\ref{fig:lc}. The shape is consistent with previously measured low-energy light curves \citep{ho2022timing,tomsick2012is}.

Due to \textit{NuSTAR} and \textit{NICER}'s limited spatial resolution, the pulsar light curve will have a constant component which cannot be separated from the trail with those data. Fortunately, the IXPE data partially resolve the pulsar from the trail as can be seen in the top panels of Fig.~\ref{fig:lc}: during the pulsar's ``off'' phase, the trail/pulsar complex's center of light moves to the northeast, farther from the pulsar. We may therefore use the spatial weights to determine the light curve constant component from the IXPE data and our \textit{Chandra} model. This is done using a simplified version of the full \texttt{LeakageLib} polarimetric fit, fitting only to the observed phases and positions of the events and ignoring polarization information. As the light curve determines the phase weights, the constant component flux is a free parameter of this model. We find a final constant component of $36\pm 12$\% the pulsed height of the light curve, as depicted in the bottom panel of Fig.~\ref{fig:lc}. 

The constant component is too faint for direct polarization measurements with these data, so we test models in which (1) it exhibits the pulsar's polarization, (2) it exhibits the trail polarization, and (3) it is unpolarized. Model 1 is preferred over 2 and 3 by an Akaike Information Criterion (AIC) difference of 19.2 and 5.2, respectively. We therefore assume henceforth that the constant component exhibits the pulsar polarization.

\subsection{Spectrum Extraction} \label{sec:spectrum}
In principle, a simultaneous spectropolarization fit to the IXPE data is possible, but it is not necessary for this observation since \textit{Chandra}'s excellent spatial and energy resolution allows better spectral characterization of these sources. We therefore extract spectra directly from the \textit{Chandra} data using \texttt{sherpa} and fix these in the IXPE polarimetric fit. The spectra have been well measured previously in the literature, so we do not discuss the results in detail here. We use a power-law model photoelectrically absorbed with \texttt{tbabs}. We first fitted for the effective hydrogen column density with the filament, trail, and pulsar data separately, giving an uncertainty-weighted mean $n_H = 9.2 \times 10^{21}$ cm$^{-2}$. We then fix this value and re-fit the \textit{Chandra} spectra, obtaining photon indices $\Gamma = 1.69(6)$, $2.15(4)$, and $1.18(5)$ for the filament, trail, and pulsar respectively. The substantial difference between the pulsar and trail indices suggests that energy weighting can help separate signal from the two sources, in addition to the spatial and phase weights.

\subsection{Turbulence Modeling} \label{sec:turbulence}
Depolarization compared to the uniform field polarization $\Pi_0$ (Eq.~\ref{eqn:max-pd}) presents an opportunity to measure the filament's magnetic turbulence. The result depends on the assumed turbulence distribution and the angle between the magnetic field and the line of sight (LoS) $\iota$. \cite{del2025polarization} presents analytical formulas for the PD assuming isotropic turbulence for any $\iota$, and assuming anisotropic turbulence for $\iota=90^\circ$. This section gives a new formula for turbulence transverse to $\bm B$ and any $\iota$, and a prescription for extracting PD constraints when $\iota$ is unknown.

Suppose that $\chi$ is the plane of sky (PoS)-projected angle between $\bm B_0$ and the magnetic field after perturbation by turbulence, $\bm B = \bm B_0 + \Delta \bm B$. We decompose the PoS projection of $\bm B$ into components $B_\parallel$ and $B_\perp$, parallel and perpendicular to the PoS projection of $\bm B_0$. The Stokes coefficients from the perturbed field are derived from standard synchrotron theory:
\begin{equation}
  I = ZB_\mathrm{PoS}^\Gamma,\quad
  Q = -\Pi_0 I \cos 2\chi,\quad
  U = -\Pi_0 I \sin 2\chi
\end{equation}
\citep{bandiera2016radio} where $B_\mathrm{PoS}^2 = B_\parallel^2 + B_\perp^2$. $Z$ depends on the observing frequency, $\Gamma$ and fundamental constants, but is independent of $B_\mathrm{PoS}$.  Its value will not come into this calculation. Taking $\Delta B$ to be distributed symmetrically around $\bm B_0$, the average value of $U$ is $\chevronsi{U} = 0$. Therefore, the PD observed from the turbulent region is
\begin{align}
  \Pi = \frac{|\chevrons{Q}|}{\chevrons{I}} = \Pi_0 \abs{-\frac{\chevrons{(B_\parallel^2 + B_\perp^2)^{\frac{\Gamma}{2}-1}(B_\parallel^2 - B_\perp^2)}}{\chevrons{(B_\parallel^2 + B_\perp^2)^{\frac{\Gamma}{2}}}}}.
  \label{eqn:pd-mid}
\end{align}
The turbulence-PD relation has been shown to depend only weakly on $\Gamma$ \citep{bandiera2016radio}, so we set $\Gamma = 2$ to achieve approximate analytical results. For Gaussian distributions of $B_\parallel$ and $B_\perp$ with mean $\mu_\parallel$ 
(recall mean $\mu_\perp = 0$) and standard deviations $\sigma_\parallel$ and $\sigma_\perp$, $\Gamma=2$ implies
\begin{equation}
  \Pi = \Pi_0\abs{-\frac{\mu_\parallel^2 + \sigma_\parallel^2 - \sigma_\perp^2}{\mu_\parallel^2 + \sigma_\parallel^2 + \sigma_\perp^2}}.
  \label{eqn:pd-mid2}
\end{equation}
For isotropic turbulence, one has $\mu_\parallel = B_0 \sin \iota$ and $\sigma_\parallel = \sigma_\perp = \Delta B / \sqrt{3}$. Eq.~\ref{eqn:pd-mid2} yields $\Pi = \Pi_0 / (1 + (2/3)\tau^2) \sin \iota$ as derived by \cite{burn1966on}. A similar result was found by \cite{del2025polarization} for $\Gamma=1.6$, further highlighting the weak dependence of the results on $\Gamma$. Solving for $\tau$, the turbulence corresponding to observed PD $\Pi$ for an isotropic distribution of $\Delta \bm B$ is
\begin{equation}
  \tau(\Pi, \iota) = \sqrt{\frac{3}{2}\frac{\Pi_0-\Pi}{\Pi}}\sin \iota,\qquad\mathrm{(isotropic)}.
  \label{eqn:pd-isot}
\end{equation}
Turbulence transverse to $\bm B_0$ is more complex. One expects $\sigma_\perp = \Delta B / \sqrt{2}$, and $\sigma_\parallel = \Delta B \cos \iota / \sqrt{2}$, giving
\begin{equation}
  \tau(\Pi, \iota) = \sqrt{2\frac{(\pm \Pi_0-\Pi)\sin^2 \iota}{2\Pi + (\pm \Pi_0-\Pi)\sin^2 \iota}} \qquad\mathrm{(transverse)}.
  \label{eqn:pd-trans}
\end{equation}
The upper sign is taken when the observed EVPA is perpendicular to $\bm B_0$, making the term in the absolute values of Eq.~\ref{eqn:pd-mid} negative. Turbulence values $\tau < \sqrt{2}$ provide this case. For larger values of $\tau$, the transverse fluctuations dominate the underlying field, making the typical realized field $\bm B$ closer to perpendicular with $\bm B_0$. The observed EVPA therefore becomes parallel to the PoS projection of $\bm B_0$, and the lower sign is taken.

For cases such as Lighthouse where $\iota$ is not known, one should marginalize over $\iota$ weighted by a prior $P_\iota$. We use $P_\iota = (\sin \iota) / 2$, corresponding to uniform prior on $\bm B_0$. We then compute $P_\tau$, the PDF of $\tau$, by drawing $\Pi$ samples from the posterior PDF returned by the fit multiplied by a prior that enforces $\Pi < \Pi_0$. We also draw $\iota$ samples from $P_\iota$. For each $(\Pi, \iota)$ pair we compute $\tau$ and $\partial \tau/\partial \Pi$ using Eqs.~\ref{eqn:pd-isot} or \ref{eqn:pd-trans}. $P_\tau$ is the histogram of these $\tau$ samples, weighted by $|\partial \tau/\partial \Pi|$ to convert to a PDF in $\tau$.

\begin{figure}
  \centering
  \includegraphics[width=\linewidth]{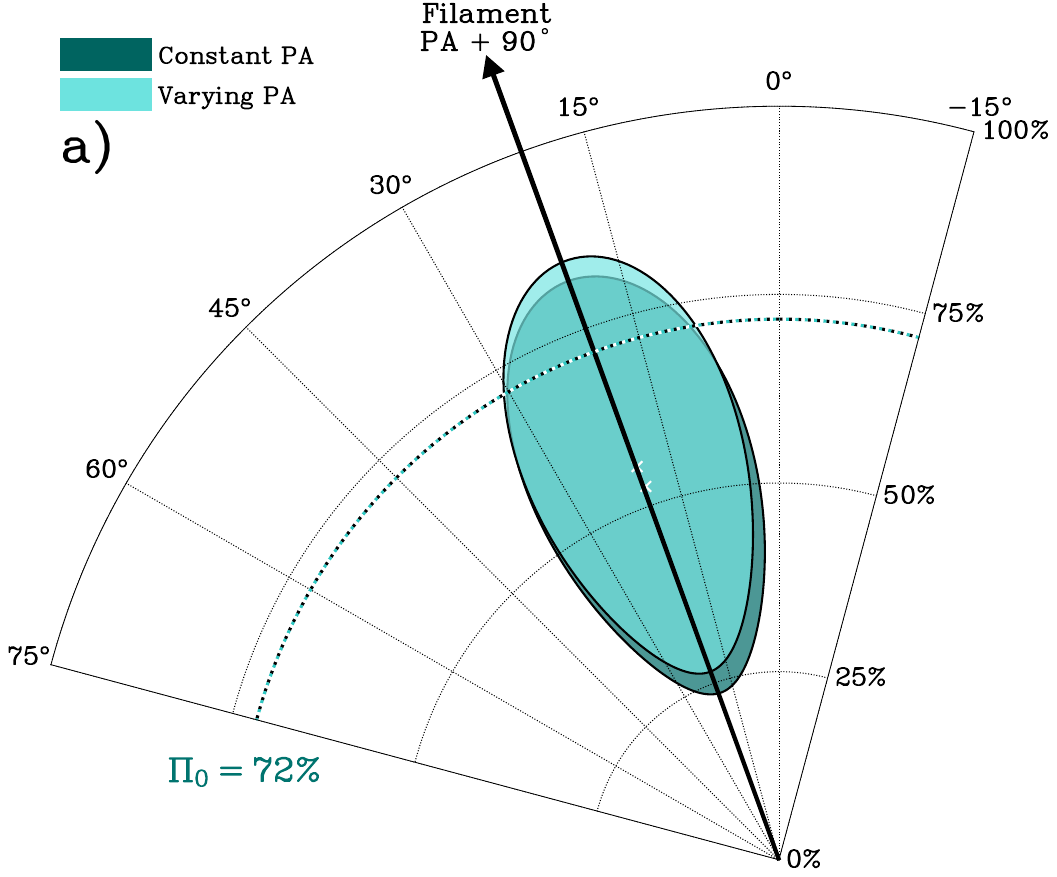}
  \includegraphics[width=\linewidth]{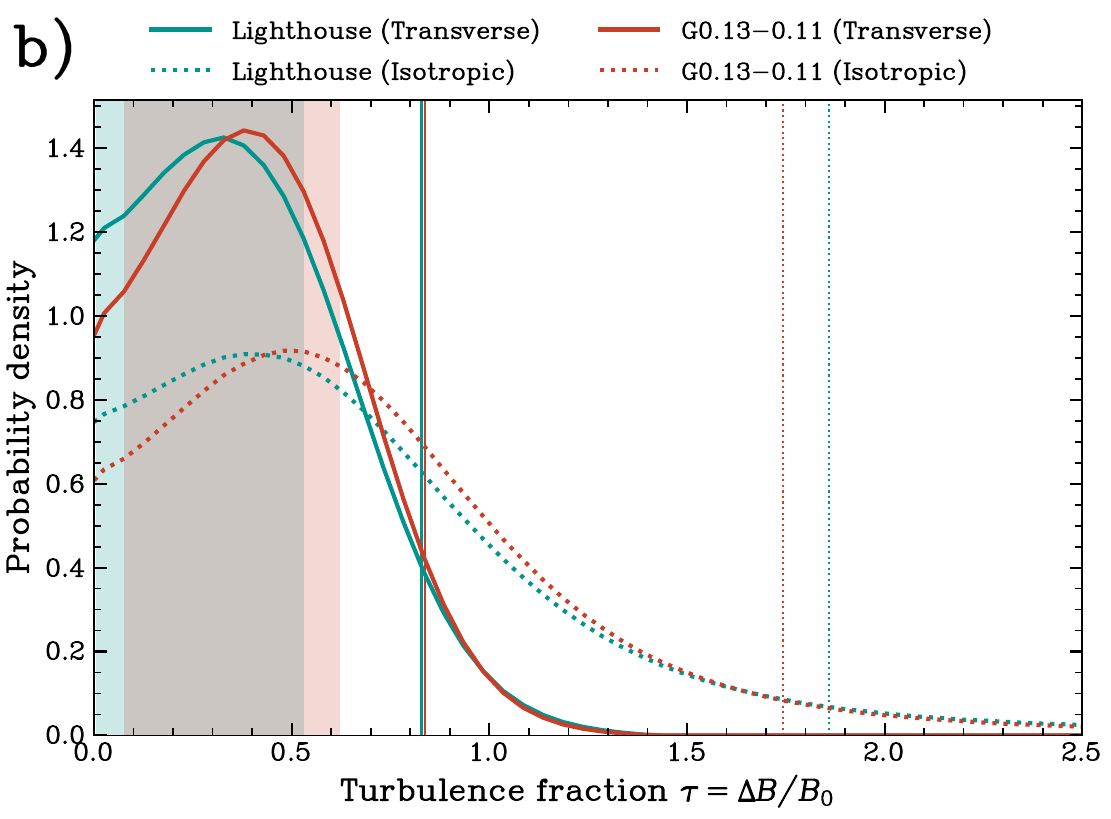}
  \caption{\textbf{a)} 68\% confidence intervals for the filament polarization. The arrow depicts the EVPA for which the magnetic field aligns with the filament, and maximum PD from Eq.~\ref{eqn:max-pd} is shown as a dotted arc.  Lighter contours show the most significant results, in which the EVPA follows an estimate of the filament normal (Fig.~\ref{fig:lighthouse}, right). Darker contours show the constant EVPA fit.  
  \textbf{b)} The turbulent magnetic field strength in the filament as constrained by the PD, for Lighthouse and the filament candidate G0.13$-$0.11. The shaded bands denote 68\% confidence intervals and vertical lines show 95\% confidence upper limits. The solid/dotted lines assume transverse/isotropic turbulence.}
  \label{fig:fil}
\end{figure}

\section{Results \& Discussion} \label{sec:results}
To study the filament polarization, we exclude all counts near the trail and pulsar and simultaneously fit the filament and background parameters. Results are discussed in \S\ref{sec:filament}. As the pulsar and trail spatially overlap, we must simultaneously fit the pulsar, trail, and background at once, using spatial, temporal, and energy weights to separate the sources. The best-fit parameters are discussed in \S\ref{sec:pulsar}/\ref{sec:trail} for the pulsar/trail.

\subsection{The Filament} \label{sec:filament}
To test the hypothesis that the magnetic field lies tangent to the filament, we had constructed an expected PoS projection of the magnetic vector field prior to receiving the IXPE data, tangent to the linear structures in the \textit{Chandra} image (Fig.~\ref{fig:lighthouse}, right). We assign the filament constant PD and EVPAs perpendicular to these vectors, rotated by a global offset $\delta \Psi$, and perform a fit. Results gave a strong polarization, with PD $\Pi =56\pm 18\%$, below the theoretical maximum of 72\% (Eq.~\ref{eqn:max-pd}). The EVPA offset $\delta \Psi = 0.4 \pm 9.3^\circ$, measured counterclockwise from north, indicates that the data are consistent with the magnetic field lines drawn in Fig.~\ref{fig:lighthouse}. The confidence intervals of this fit are shown as an offset from the black arrow in Fig.~\ref{fig:fil}a. With $p$-value of 0.8\%, this result surpasses the minimum detectable polarization to 99\% confidence (MDP$_{99}$) threshold, but without DU2 it is not $3\sigma$ significant.

As a check, we also perform a simpler fit, with constant PD and EVPA across the filament. This fit delivered $\Psi=20\pm 10^\circ$. The blue line in Fig.~\ref{fig:lighthouse} and lighter contours of Fig.~\ref{fig:fil}a show this result, which is consistent with the filament angle. The PD has slightly decreased to $\Pi=53 \pm 18\%$ and the significance to $p=1.2\%$, presumably because of depolarization inherent in averaging over the varying filament orientation. We therefore use the varying EVPA model as our primary model. However, the significance increase does not suffice to confidently claim a variable EVPA; that would require substantially longer exposures.

Fig.~\ref{fig:fil}b presents our constraints on the turbulent field strength, using the methods of \S\ref{sec:turbulence}. Most filament models put $\bm B_0$ parallel to the filament and generate turbulence through a streaming instability, which is usually transverse to the underlying field \citep[e.g. the non-resonant streaming instability, NRSI;][]{bell2004turbulent}. As an EVPA parallel to the filament is excluded to $>99\%$ confidence, the transverse turbulence strength must be small (the upper sign of Eq.~\ref{eqn:pd-trans}; see surrounding text). The IXPE PD gives $\tau = \Delta B / B_0 = 33^{+20}_{-32}\%$ with a 95\% confidence upper limit of $\tau < 0.83$, averaged over the filament. As an alternative, the 95\% upper limit assuming isotropic turbulence is $\tau < 1.84$ (dotted lines in Fig.~\ref{fig:fil}b). We argue that the transverse model has the best physical motivation, but both models exclude a case where the NRSI saturates throughout the filament, at e.g.~$\tau\sim 5$. Fig.~\ref{fig:fil}b also shows turbulence constraints for G0.13$-$0.11 using IXPE polarizations presented by \cite{churazov2024pulsar}. They similarly exclude strong turbulence. However, lacking pulsations from its low velocity point source \citep[$<$160 km s$^{-1}$ transverse velocity,][]{dinsmore2026chandra}, this object is not a confirmed pulsar filament.

\subsection{The Pulsar} \label{sec:pulsar}
\begin{figure}
  \centering
  \includegraphics[width=\linewidth]{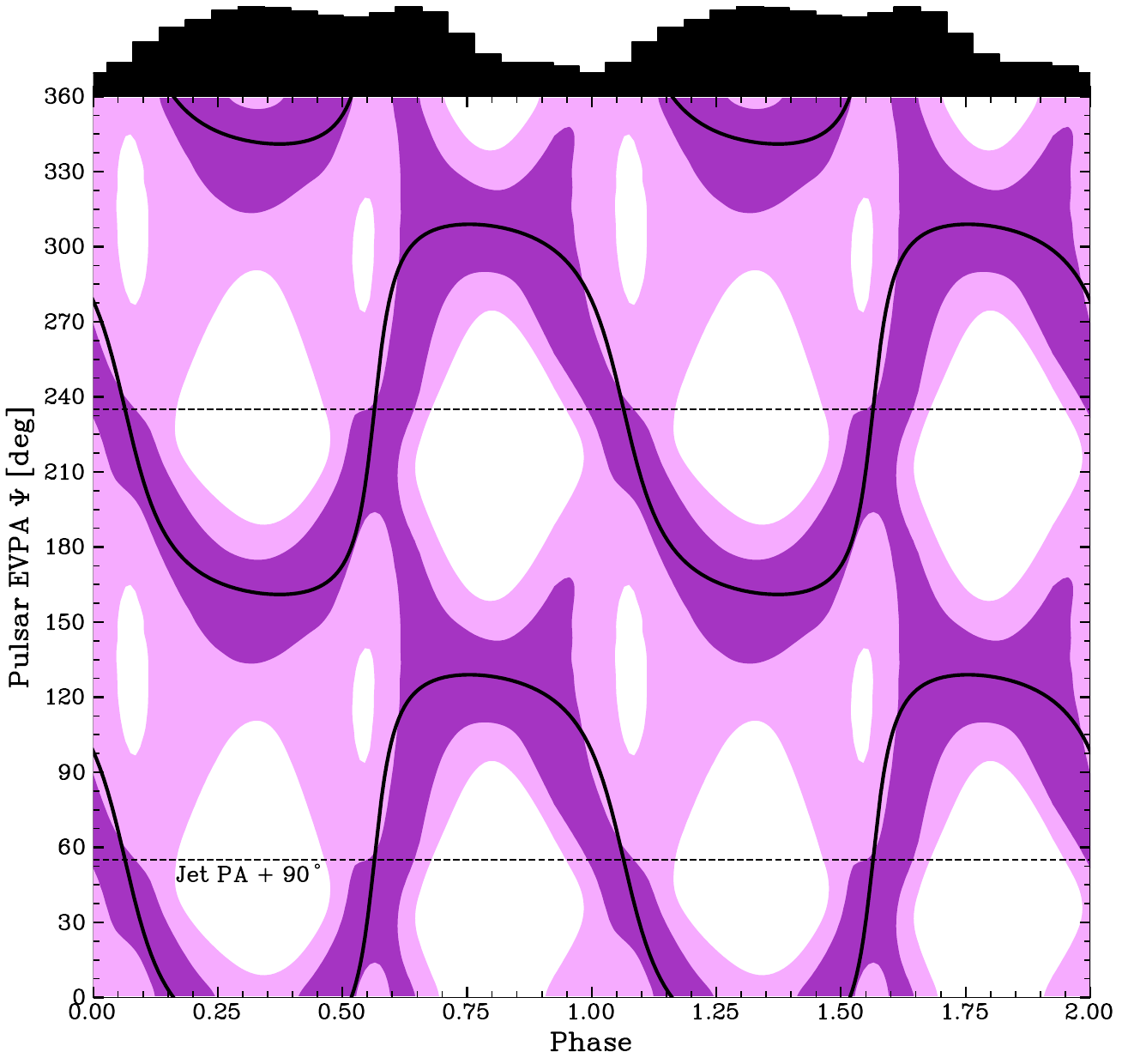}
  \caption{1$\sigma$ and 2$\sigma$ confidence intervals for the pulsar's RVM sweep as a function of phase. The IXPE light curve is presented over the plot for comparison. For clarity, EVPA and phase are plotted over twice the detectable range (i.e.~$0-2$ for phase and $0-360^\circ$ for EVPA).}
  \label{fig:pulsar}
\end{figure}
We model the pulsar polarization using the rotating vector model \citep[RVM;][]{radhaskrishnan1969magnetic}, which describes the pulsar EVPA as emitted by a magnetic axis rotating steadily around the spin axis. Its four parameters are the spin axis-magnetic axis angle $\alpha$, the spin axis-LoS angle $\zeta$, the phase of closest approach between the magnetic axis and LoS $\phi_0$, and the EVPA at $\phi_0$ $\Psi$. Since EVPA is perpendicular to the projected field for synchrotron emission, we set $\Psi$ perpendicular to the position angle (PA) of the spin axis, which is inferred to be $-35^\circ$ from the narrow (polar jet-like) outflow seen near the pulsar in \textit{Chandra} data \citep{klingler2023nustar}. This leaves three RVM parameters, to which we must add a fourth representing pulsar average PD because our unbinned fit method requires a PD prediction.

The best-fit pulsar parameters give a high PD of $59\pm 16$\%, approaching the synchrotron limit of $\Pi_0 = 64\%$. Such high X-ray PD has also been inferred by IXPE for PSR B0540$-$69, which has a similar pulse profile \citep{xie2024first}. The RVM parameters are $\phi_0 = 0.565\pm 0.043$, $\alpha = 73\pm 22^\circ$, and $\zeta = 83 \pm 37^\circ$, with correlations between the parameters. These uncertainty estimates assume Gaussian posteriors, though the true posteriors are likely more complex. Fig.~\ref{fig:pulsar} shows 1$\sigma$ and 2$\sigma$ uncertainties on the sweeping pulsar EVPA, accounting for the full non-Gaussian probability distribution. The angle between the magnetic field and LoS at closest approach is consistent with large angles ($\beta = \zeta - \alpha = 11 \pm 36^\circ$) as expected given the lack of pulsar radio emission. The pulsar/trail fit is 3.2$\sigma$ significant compared to one in which both sources are unpolarized, rendering this a secure detection of polarization in the complex. However our best-fit is significant to only 2.6$\sigma$ compared to the case of a polarized trail and unpolarized pulsar, corresponding to $p=1\%$.

\subsection{The Trail} \label{sec:trail}
\begin{figure}
    \centering
    \includegraphics[width=\linewidth]{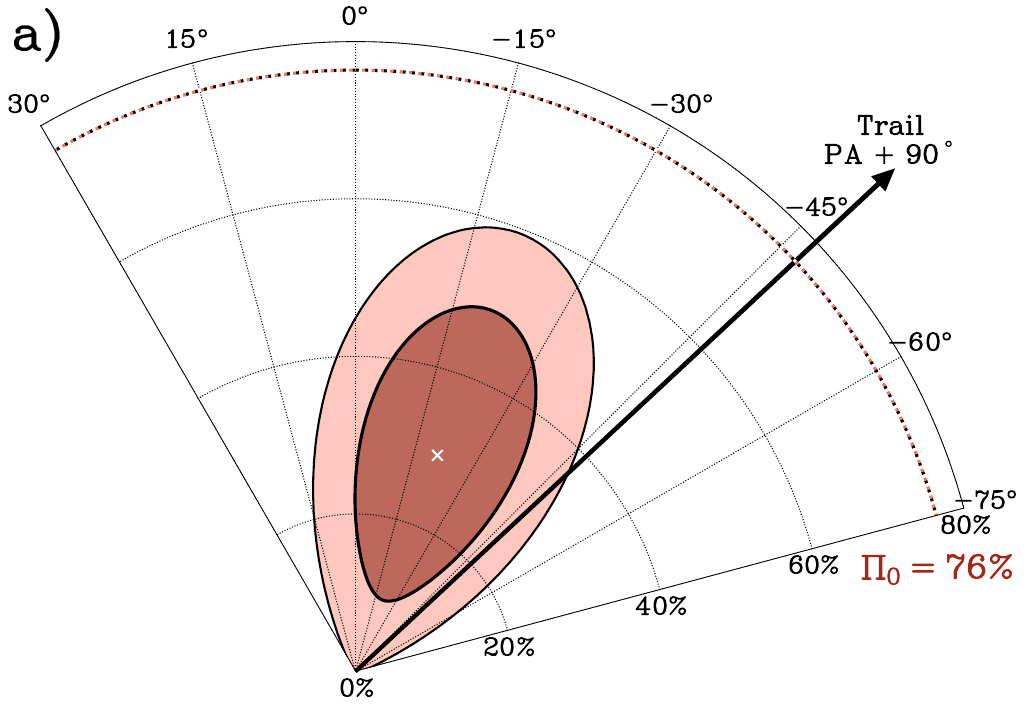}
    \includegraphics[width=\linewidth]{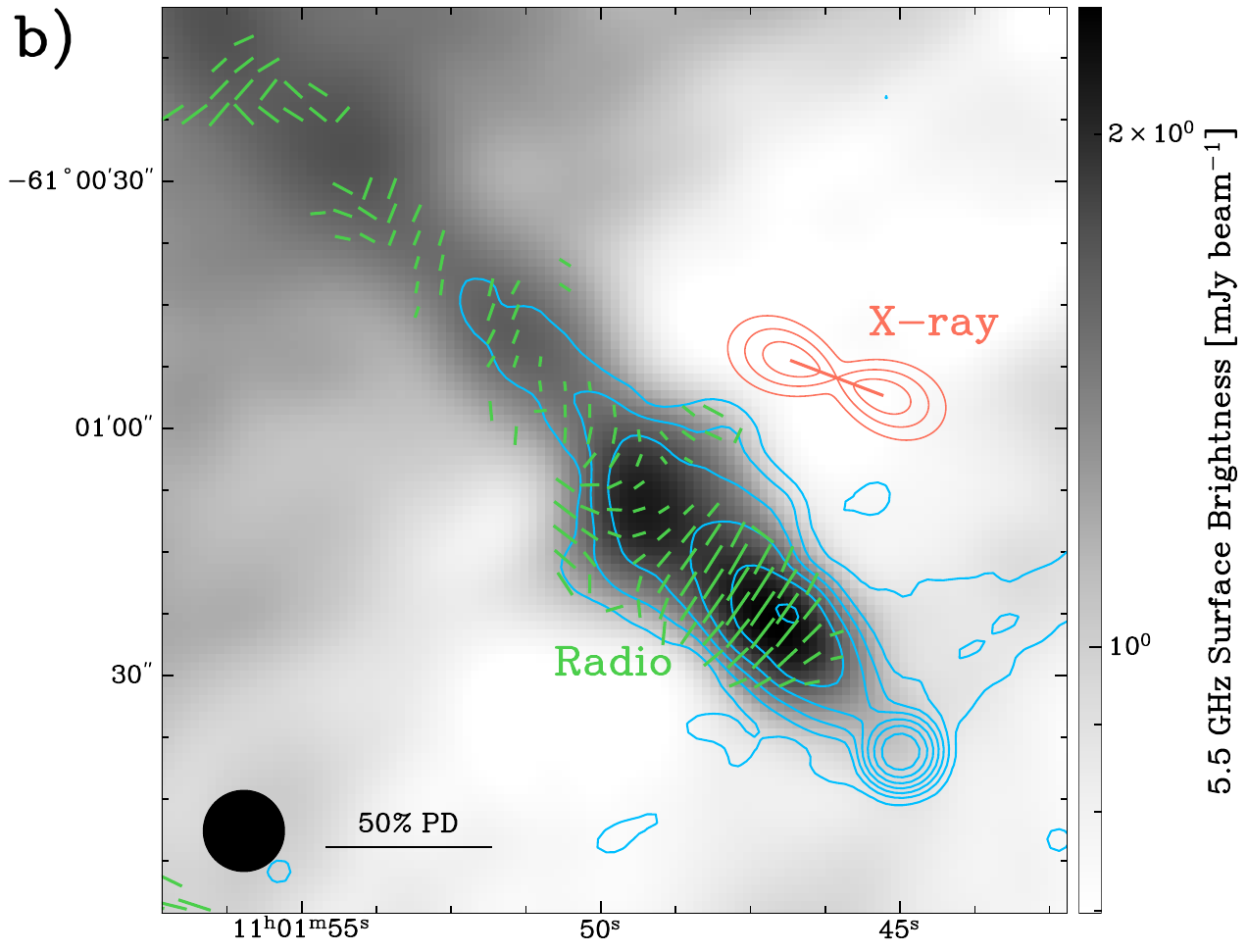}
    \caption{\textbf{a)} Polarization results for the trail, with the maximum PD from Eq.~\ref{eqn:max-pd} shown as a dotted arc. The arrow denotes the EVPA for which the magnetic field aligns with the trail. \textbf{b)} 5.5 GHz ATCA intensity map of the Lighthouse trail, with magnetic field vectors inferred from the radio polarization overlaid in green. The IXPE X-ray trail magnetic field vector is also shown in orange and is almost orthogonal to the radio. \textit{Chandra} contours are shown in blue, and the simulated radio beam in the lower left. There is no radio filament or pulsar emission.}
    \label{fig:radio}
\end{figure}

In the simultaneous pulsar and trail fit, the trail component has an EVPA $\Psi=-24 \pm 12^\circ$ consistent with magnetic fields parallel to the trail axis. The trail polarization degree of $\Pi = 26\pm 11\%$ is well below the synchrotron limit of $\Pi_0 = 76\%$, indicating strong turbulence as expected. Compared to the case of a polarized pulsar and unpolarized trail, our result is significant to 1.9$\sigma$.

\begin{table*}
    \centering
    \hspace{-2.4cm}
    \begin{tabular}{ccccccc}
        \hline\hline
        Observation & Array & Maximum & Integration & No.\ of & Center & Bandwidth \\
        Date & configuration & baseline [m] & time [hr] & pointings & frequency [MHz] & [MHz] \\ \hline
        2015 Dec 3 & 1.5A & 4469 & 10.5 & 4 & 5500, 9000 & 2049\\
        2015 Dec 11 & 750C & 5020  & 10.25 & 4 & 5500, 9000 & 2049\\
        \hline \hline
    \end{tabular}
    \caption{Parameters of the ATCA Lighthouse observation.}
    \label{t:obs}      
\end{table*}

We can compare this result with Australia Telescope Compact Array (ATCA) radio polarization observations of the trail. These were made on 2015 Dec 3 and 11 in the 1.5A and 750C array configurations respectively, with a total of 20\,hr on-source time. Data were taken simultaneously at the 6 and 3\,cm bands centering at 5.5 and 9\,GHz, respectively, with bandwidth of 2\,GHz each. The observation parameters are listed in Tab.~\ref{t:obs}. We performed the data analysis using the MIRIAD package \citep{Sault:1995}. After the standard flagging and calibration procedure, we employed the peeling technique \citep{hughes:2007} to mitigate strong sidelobes due to a bright point source in the north. Radio images were formed using the multi-frequency synthesis technique with the \citet{bri95} weighting parameter of \texttt{robust=0.5}. They were then deconvolved with the maximum entropy method (MEM) and restored with a circular beam of FWHM $10\arcsec$. The rms noise level of the 6\,cm total intensity map is $\sim60$\ujb, significantly higher than the theoretical value of 18\ujb, due to sidelobes from the bright source in the north. We generate polarization intensity (PI) and position angle (PA) maps using the cleaned Stokes Q and U images. We estimate the foreground Faraday rotation by comparing between the 6 and 3\,cm PA maps. The radio intensity and magnetic field vectors are shown in Fig.~\ref{fig:radio}b, together with contours from the \textit{Chandra} X-ray image and the IXPE average trail magnetic axis.

\begin{table*}
  \centering
  \hspace{-2.5cm}
  \begin{tabular}{r|ccc|ccc}
  \hline \hline
  \hspace{0.1cm} & & NN & & & Moments & \\
  \hspace{0.1cm} & $Q$ & $U$ & Sig.  & $Q$ & $U$ & Sig. \\\hline
  Filament \texttt{LeakageLib} & $0.41 \pm \textbf{0.18}$ & $0.33 \pm \textbf{0.19}$ & 2.4$\sigma$ &$ 0.48 \pm 0.22$ & $0.40 \pm 0.21$ & 2.4$\sigma$\\
  Filament \texttt{PCUBE} & $0.40 \pm 0.35$ & $0.16 \pm 0.35$ & 0.7$\sigma$ & $0.110 \pm 0.33$ & $0.38 \pm 0.33$ & 0.7$\sigma$ \\ \hline
  PSR+trail \texttt{LeakageLib} & $0.188 \pm \textbf{0.083}$ & $-0.208 \pm \textbf{0.083}$ & 2.9$\sigma$ & $0.284 \pm 0.098$ & $-0.240 \pm 0.098$ & 3.4$\sigma$ \\ 
  PSR+trail \texttt{PCUBE} & $0.13 \pm 0.14$ & $-0.15 \pm 0.14$ & 0.9$\sigma$ & $0.21 \pm 0.13$ & $-0.11 \pm 0.13$ & 1.4$\sigma$ \\
  \hline \hline 
  \end{tabular}
  \caption{Polarization degrees, angles, and significances for the filament and pulsar+trail obtained through alternative analyses. This table shows fits assuming constant polarizations in both regions to allow comparison with the \texttt{PCUBE} aperture polarimetry method; our main results use the more complex, filament-following model. Our results use advanced weights provided by a \texttt{LeakageLib} analysis with NN reconstruction, which give decreased uncertainties.}
  \label{tab:analyses}
\end{table*}

The radio fields are largely perpendicular to the trail axis, which is contrary to IXPE's result of parallel fields. Indeed the flux-weighted average radio PD in the \textit{Chandra} ``trail'' region is $\Pi = 2\%$, $\Psi=56^\circ$, and the maximum PD is $\Pi = 12\%$ with $\Psi=55^\circ$; these polarizations are inconsistent with the IXPE finding to 2.1 and 2.9$\sigma$. One possible explanation is that the radio- and X-ray emitting electrons occupy trail regions with different field geometries. In relativistic MHD simulations, \cite{olmi2019full} showed that for isotropic pulsar winds, the shock trail can exhibit a layered structure, with an outer shell of strong magnetic field parallel to the trail and an inner core with a more turbulent field, largely perpendicular to the trail. The radio image seems to support this interpretation, as it contains a long narrow trail extending farther from the pulsar than any X-ray counterpart. This may be the remnant perpendicular-field core after the X-ray electrons have cooled and/or escaped to the ISM. Another possibility is that internal Faraday rotation from fluctuating densities of cold electrons in the trail lowers the radio PD and affects the EVPA; such depolarization is seen in other pulsar wind nebulae \citep[e.g.][]{lai2026high}.

Recent 0.9$-$1.7\,GHz MeerKAT observations have detected several thin streaks transverse to the trail with no X-ray counterpart \citep{martin2026radio}. These are plausibly due to a low energy electron population escaping to ISM structures generated by earlier filament injection epochs; additional study might then shed light on the filament formation.

\section{Alternative Analyses} \label{sec:variations}

It is useful to compare our results to other analysis methods to confirm the primary findings. Our results use \texttt{LeakageLib} to leverage spatial, phase, energy, and particle weights with a maximum likelihood estimator. A simpler alternative is the aperture polarimetry \texttt{PCUBE} analysis pipeline provided by \texttt{IXPEobssim} \citep{baldini2022ixpeobssim}, which does not use these weights. Another analysis choice is the method of data reconstruction. We used an NN method which delivers more accurate results and estimates modulation factors for each events. An alternative is the mission-standard Moments method, which only estimates modulation factor as a function of event energy. This section demonstrates that the four methods are consistent and the \texttt{LeakageLib}+NN method is the most precise.

Given events with EVPA $\psi_i$ and therefore Stokes coefficients $q_i = \cos 2\psi_i$ and $u_i = \sin 2\psi_i$, \texttt{PCUBE} estimates source Stokes coefficients using $Q = 2\chevronsi{q_i / \mu_i}$ and  $U = 2\chevronsi{u_i / \mu_i}$, where $\mu_i$ is the event modulation factor and $\chevrons{-}$ indicate averaging over events within the aperture. Linearly propagated from these equations, the uncertainties are
\begin{equation}
    \sigma_Q^2 = \frac{1}{N}\chevrons{\frac{2}{\mu_i^2} - Q^2}, \qquad
    \sigma_U^2 = \frac{1}{N}\chevrons{\frac{2}{\mu_i^2} - U^2},
    \label{eqn:pcube}
\end{equation}
although the orignal \texttt{PCUBE} code adopts another uncertainty formula, which under-estimates errors. For this analysis, we cut background particles with the method designed by \cite{dimarco2023handling}. The \texttt{PCUBE} results are unweighted and were background subtracted using a large region to the south of the filament.

The top section of Tab.~\ref{tab:analyses} presents results for the filament, using Eq.~\ref{eqn:pcube} for \texttt{PCUBE} uncertainties. The bottom section presents results for the combined polarization of the pulsar+trail complex. From the \textit{Chandra} image, approximately $2/3$ of the non-background counts are emitted by the trail with the remainder emitted by the pulsar. The EVPAs and PDs are mutually consistent between the methods, and the \texttt{PCUBE} results concur with the main findings of our analysis, including a strongly polarized filament and trail EVPA perpendicular to the radio result. But the advanced weights of \texttt{LeakageLib} and NN processing are necessary to make these results significant. In cases where the Moments result delivers higher $Q$ or $U$, the nominal significance can be slightly larger than the NN value. But since the Moments uncertainties are always larger, this should be interpreted as a statistical fluctuation and not a signal/noise improvement.

\section{Conclusions} \label{sec:conclusion}
We have detected X-ray polarization in Lighthouse's filament to $>99\%$ confidence, and in the pulsar/trail complex to $>3\sigma$ confidence. This marks the first X-ray polarization results for an X-ray trail and for a filament confirmed to emanate from a pulsar. The following results are significant to 99\% confidence: In the filament, the magnetic field is tangent to the X-ray structure and only weakly turbulent. In the trail, the X-ray EVPA is misaligned with---indeed nearly orthogonal to---the radio EVPA. The pulsar is highly polarized and well-modeled by the rotating vector model (RVM).

The Lighthouse filament's high polarization and EVPA perpendicular to the X-ray structure echoes IXPE results for filament candidate G0.13$-$0.11 \citep{churazov2024pulsar}. Our large PD detection rejects isotropic turbulence with $\Delta B / B > 1.84$ and transverse turbulence with $\Delta B / B > 0.83$ to 95\% confidence. Our detection of EVPA perpendicular to the filament rejects transverse turbulence with $\Delta B / B > \sqrt{2}$ to $>99\%$ confidence because such strong transverse turbulence would rotate the EVPA. These results are in conflict with strong turbulence models \citep[e.g.][the latter of which predicts $\Delta B / B = 8.7 B_{3\, \mathrm{\mu G}}^{-1}$ in Lighthouse]{bykov2017pulsar,olmi2024nature}. Models where the ambient ISM field is only weakly perturbed seem preferable; one such picture is described in \cite{dinsmore2026physical}. For the trail, the differing radio and X-ray EVPAs indicate that the X-ray and radio-emitting particles may illuminate different regions of a complex, turbulent magnetic field structure. Though the pulsar's polarization is significant, the RVM results are sufficiently uncertain to preclude firm physical conclusions.

Our results give $1.8\times$ and $1.6\times$ smaller PD errors compared to the aperture polarimetry \texttt{PCUBE} method for the filament and pulsar respectively, due to new analysis techniques explored in several recent studies. The weights are particularly effective for the filament, which is the fainter structure. This highlights the importance of using advanced weights when analyzing low surface brightness observations, as weights help separate the (polarized) background events. Approximately 2.2 additional megaseconds of IXPE time would have been required to achieve the same $1.8\times$ improvement with data alone.

Improved detections should, however, be sought. The polarizations of the pulsar and trail as individual components hover at $\sim2\sigma$ confidence, and the significances of the pulsar polarization sweep and EVPA / PD variation within the filament are similarly modest. However, with advanced weighting analysis, additional IXPE exposure of this unique object can improve constraints on the magnetic field structure in the trail and filament, and on the pulsar polarization.

\begin{acknowledgments}
This work was funded in part by NASA grant 80NSSC25K0278 administered by the Goddard Space Flight Center and by grant GO4-25037B administered by the Smithsonian Astrophysical Observatory (SAO). C.-Y. N. and S. Zhang are supported by GRF grants of the Hong Kong Government under HKU 17304524.
\end{acknowledgments}

\vspace{5mm}
\facilities{IXPE, CXO, ATCA}
\software{LeakageLib \citep{dinsmore2024leakagelib}, HEAsoft \citep{nasa2014HEAsoft}, IXPEobssim \citep{baldini2022ixpeobssim}, PINT \citep{luo2021pint}}


\bibliography{bib}{}
\bibliographystyle{aasjournal}

\end{document}